\documentclass[aps,prd,twocolumn,draft,nofootinbib]{revtex4-1}
\usepackage{amsmath}
\usepackage{amsthm}
\usepackage{amsfonts}

\newcommand{\vnab}{\vec{\nabla}}

\begin{document}

\title{Singular Hamiltonians in models with spontaneous Lorentz
  symmetry breaking}

\author{Michael D. Seifert}
\affiliation{Dept.\ of Physics, Astronomy, and Geophysics, Connecticut
  College \\ 270 Mohegan Avenue, New London, CT 06320, USA} 
\email{mseifer1@conncoll.edu}
\date{\today}

\begin{abstract}
  Many current models which ``violate Lorentz symmetry'' do so via a
  vector or tensor field which takes on a vacuum expectation value,
  thereby spontaneously breaking the underlying Lorentz symmetry of
  the Lagrangian.  One common way to construct such a model is to
  posit a smooth potential for this field; the natural low-energy
  solution of such a model would then be excepted to have the tensor
  field near the minimum of its potential.  It is shown in this work
  that some such models, while appearing well-posed at the level of
  the Lagrangian, have a Hamiltonian which is singular on the vacuum
  manifold and are therefore ill-posed.  I illustrate this pathology
  for an antisymmetric rank-2 tensor field, and find sufficient
  conditions under which this pathology occurs for more general field
  theories. 
\end{abstract}

\maketitle

\section{Introduction}

The prospect of finding new physics via Lorentz symmetry violation has
been of significant interest over the past couple of decades.  In many
such models, Lorentz symmetry is broken spontaneously: one postulates
the existence of a new fundamental field that is not a Lorentz scalar,
and assigns dynamics to this field that obey Lorentz symmetry but lead
it to take on a non-zero ``vacuum expectation value''.  The existence
of this non-zero Lorentz vector or tensor field then provides a
preferred geometric structure in spacetime.

In effect, such a field would provide a ``hook'' upon which one can
``hang'' frame-dependent effects.  In the presence of a Yukawa-like
couplings between this new field and conventional matter fields, the
results of experiments would depend on the relative orientation in
spacetime of the observer's four-velocity and the new field, leading
to frame-dependent effects.  Such a field is frequently called a
``Lorentz-violating'' (LV) field, though this is something of a
misnomer; the postulated field would still transform between frames
via the standard Lorentz transformation laws.  The dynamics of such
fields in flat spacetime have been studied in their own right
\cite{Kostelecky2004, Kostelecky2009, Seifert2009, Altschul2010}, and
similar models have also been developed in the context of curved
spacetime as possible modifications to general relativity
\cite{Jacobson2001, Bekenstein2004, Moffat2006}.

Many (though not all) of the above-cited models share two features.
First, they accomplish the spontaneous breaking of Lorentz symmetry by
assigning a potential energy $V(\Psi^{\cdots})$ to some Lorentz tensor
$\Psi^{\cdots}$.  This potential is constructed in a ``Higgs-like''
way, so that it is minimized on a \emph{vacuum manifold} in field
space.  Since the field $\Psi^{\cdots}$ is a Lorentz tensor, but we
want the underlying equations of motion to obey Lorentz symmetry, we
have to construct the potential out of one or more Lorentz scalars
that are dependent on $\Psi^{\cdots}$.  The vacuum manifold will be
determined by a set of conditions on these Lorentz scalars.  For
example, if we wish to construct a model in which a Lorentz vector
field $A^a$ spontaneously breaks Lorentz symmetry, it is not hard to
see that the potential $V(A^a)$ must be some function of the
four-vector norm $A_a A^a$, and the vacuum manifold will be the set of
all four-vectors of a particular norm.

Second, the kinetic terms in the Lagrangians for these theories must
be constructed with care.  Assuming that the underlying Lagrangian is
second-order and gives rise to second-order equations of motion, the
kinetic terms will have to involve a contraction of the spacetime
derivative $\nabla_a \Psi_{b_1 \cdots b_n}$ with itself.  However, one
cannot usually simply write down a kinetic term of the form $\nabla_a
\Psi_{b_1 \cdots b_n} \nabla^a \Psi^{b_1 \cdots b_n}$ and call it a
day.  Since the Minkowski metric is indefinite, such a kinetic term
will lead to terms with the ``wrong sign'' of the kinetic energy in
the Hamiltonian, which will generically lead to instabilities in the
classical solutions.  Instead, the kinetic term is usually constructed
out of more complicated combinations of the field derivatives, such
that the problematic terms in the kinetic energy do not appear (via
cancellations between various contractions).  Again returning to the
case of a four-vector field for illustration, it is not hard to see
that a kinetic term of the form $\frac{1}{2} \nabla_a A_b \nabla^a
A^b$ will include terms of the form $\frac{1}{2} (\dot{A}_0)^2 -
\dot{\vec{A}}^2$, which is indefinite.  However, the time derivatives
that arise from ``Maxwell'' kinetic term $-\frac{1}{4} F_{ab} F^{ab}$
(with $F_{ab} = 2 \nabla_{[a} A_{b]}$) are all of the same sign,
evading this problem.

The price one pays for this good behaviour, however, is that the time
derivatives of some fields do not appear at all in the Lagrangian.
This leads to the model being \emph{constrained}: certain canonical
field momenta must vanish identically, and so one cannot freely
specify both the values of the fields and their time derivatives at
some initial moment $t_0$.

The purpose of this work is to illustrate and explore a potential
incompatibility between these two features.  Specifically, a model
which contains a constrained field with a vacuum manifold, while
appearing well-posed at the level of the Lagrangian, may in fact have
a Hamiltonian that becomes singular on the vacuum manifold.  This
raises serious doubts about the viability of such a theory: it
predicts that the field, if perturbed slightly from its vacuum
manifold, could not smoothly evolve back to the vacuum manifold.  This
conflict does not occur in all Lorentz-violating field models, but
instead seems to arise when the ``number of constraints'' exceeds the
codimension of the vacuum manifold in field space.

The work is structured as follows.  A brief summary of the techniques
of Hamiltonian field theory pertinent to this result is presented in
Section \ref{HFTsec}.  Section \ref{AStenssec} contains an
illustration of how a singular Hamiltonian can arise in a
Lorentz-violating field theory, by examining the dynamics of a
rank-two antisymmetric tensor field $B_{ab}$ in a Lagrangian that is
designed to spontaneously break Lorentz symmetry.  Finally, Section
\ref{GFT.sec} discusses how this pathology could arise in a general
field theory.

Throughout this work, I will use units in which $\hbar = c = 1$; the
metric signature will be $({}-{}+{}+{}+{})$.  Roman indices $a, b, c,
\dots$ will be used to denote spacetime tensor indices; $i, j, k,
\dots$ will be used to denote spatial indices, where necessary.  Greek
indices $\alpha, \beta, \gamma, ...$ will generally only be used to
denote indices in field space.  All expressions involving repeated
indices (either tensor indices or field space indices) can be assumed
to obey the Einstein summation convention, unless explicitly stated
otherwise; in particular, for a significant fraction of Section
\ref{gencase.sec} and in the Appendix, the field-space summations will
be written out explicitly.  While I will be working exclusively in the
realm of flat spacetime, I will still use the symbol $\nabla_a$ (or
$\nabla_i$) to denote spacetime (or spatial) derivatives; the symbol
$\partial$ will be reserved for partial derivatives of functions of
fields (such as the field potential energy or the Lagrangian density)
with respect to their arguments.  The symbol $\delta$ will generally
denote either variations of functionals or functional derivatives.

\section{Hamiltonian field theory \label{HFTsec}}
In classical mechanics, the construction of a Hamiltonian from a
Lagrangian $L(q_\alpha,\dot{q}_\alpha)$ is a relatively
straightforward process.  One performs a Legendre transform on the
Lagrangian, by defining the conjugate momenta $p_\alpha = \partial
L/\partial \dot{q}_\alpha$; inverting these relationships to write the
velocities $\dot{q}_\alpha$ in terms of the coordinates $q_\alpha$ and
the momenta $p_\alpha$; and finally writing the Hamiltonian as
$H(q_\alpha, p_\alpha) = p_\alpha \dot{q}_\alpha - L(q_\alpha,
\dot{q}_\alpha)$, now viewing $\dot{q}_\alpha$ as a function of
$q_\alpha$ and $p_\alpha$.  One can then find the evolution of the
coordinates and momenta via Hamilton's equations, or via the Poisson
bracket:
\begin{equation}
  \frac{d f}{dt} = \{ f, H \} \label{hamtimeev}
\end{equation}
where for any quantities $f$ and $g$ we have
\begin{equation}
  \{ f, g\} \equiv \sum_\alpha \frac{\partial f}{\partial q_\alpha}
  \frac{\partial g}{\partial p_\alpha} - \frac{\partial g}{\partial q_\alpha}
  \frac{\partial f}{\partial p_\alpha}.
\end{equation}

In field theory for some set of fields $\psi_\alpha$, one would like
to follow the same procedure starting from a Lagrange density
$\mathcal{L}(\psi_\alpha, \dot{\psi}_\alpha, \nabla_i \psi_\alpha)$.
However, an important difference can arise when one attempts to
perform the Legendre transform.  It can happen that due to the
structure of the kinetic terms, the field velocities cannot all be
written in terms of the field momenta.  In general, this implies that
one or more of the equations of motion for the field are actually
constraint equations, not evolution equations.

One can, however, still attempt to construct a Hamiltonian that
generates the time-evolution of the system (in the sense of
\eqref{hamtimeev}) via a construction due to Dirac and Bergmann
\cite{Dirac1964}.  A brief description of the method can also be found
in a paper by Isenberg and Nester \cite{Isenberg1977}, and the
notation used here will largely follow the notation in that work.  The
construction proceeds as follows:

\begin{itemize}
\item Define the field momenta via the natural generalization:
  \begin{equation}
  \pi_\alpha = \frac{\partial \mathcal{L}}{\partial
    \dot{\psi}_\alpha}. \label{genpidef}
  \end{equation}
  One can then attempt to invert these relationships to find
  $\dot{\psi}_\alpha$ as a function of $\pi_\alpha$, $\psi_\alpha$,
  and their derivatives.  However, it may occur that certain equations
  (or combinations of equations) in \eqref{genpidef} do not contain
  any velocities.  These equations must be thought of as constraining
  the initial data, and can be written in the form
  \begin{equation}
    \Phi_I (\psi_\alpha, \vnab \psi_\alpha, \pi_\alpha) =
    0 \label{genconstraints} 
  \end{equation}
  for $I = 1, 2, \dots, M$.  The functions $\Phi_I$ are known as the
  \emph{primary constraints}.

\item Construct the \emph{base Hamiltonian density} via a Legendre
  transform on the Lagrange density:
  \begin{equation}
    \mathcal{H}_0 \equiv \pi_\alpha \dot{\psi}^\alpha - \mathcal{L}.
  \end{equation}
  It can be shown that all of the velocities $\dot{\psi}^\alpha$ will
  vanish in this process.  However, the evolution generated by this
  base Hamiltonian $H_0 = \int d^3 x \mathcal{H}_0$ will not, in
  general, preserve the constraints \eqref{genconstraints}.  To obtain
  a Hamiltonian which preserves the constraints, one must ``augment''
  the base Hamiltonian density by adding the constraints to it, each
  multiplied by an as-yet undetermined Lagrange multiplier $u_I$:
  \begin{equation} \label{augHamdef}
    \mathcal{H}_A \equiv \mathcal{H}_0 + u_I \Phi_I.
  \end{equation}
  This latter quantity is the \emph{augmented Hamiltonian density}.

\item If the constraints are to be preserved by the augmented
  Hamiltonian $H_A = \int d^3 x \mathcal{H}_A$, it must be the case
  that $\{ \Phi_I, H_A \} = 0$ for each constraint.  If this Poisson
  bracket does not vanish identically, this will yield a
  \emph{secondary constraint} $\Psi_I = 0$, which the initial data
  must also obey.  Similarly, this secondary constraint must also be
  conserved, so we demand $\{ \Psi_I, H_A \} = 0$ as well; this will
  then generate further secondary constraints.  The requirement that
  the constraints be preserved may determine some or all of the
  hitherto undetermined Lagrange multipliers, in which case we can
  replace them in the augmented Hamiltonian with an expression written
  solely in terms of the fields and the momenta.  It is also
  conceivable that we may find an inconsistent model (i.e., one of the
  constraint equations cannot be preserved under the time evolution
  generated by $H_A$).

\end{itemize}
When the dust settles, one is left with a Hamiltonian $H_A$ which
generates the time evolution of the fields.  We can use this
Hamiltonian to count (in a simplified way) the number of degrees of
freedom of the system;  specifically,
\begin{multline} 
  \label{generalcount}
  N_{dof} = \frac{1}{2} \left[ \genfrac(){0pt}{0}{\text{no.\
        of}}{\text{fields}} +
    {\text{no.\ of} \choose \text{momenta}} \right. \\
  \left. {} - {\text{no.\ of} \choose \text{constraints}} -
    {\text{no.\ of undetermined} \choose \text{Lagrange multipliers}}
  \right].
\end{multline}
The undetermined Lagrange multipliers that remain after this process
are associated with gauge degrees of freedom, and so are unphysical.
In the present work, however, the models under consideration will have
all of their Lagrange multipliers determined, and (since the number of
fields and momenta will be the same) we will have 
\begin{equation}
  N_{dof} = {\text{no.\ of} \choose \text{fields}} - \frac{1}{2}
  {\text{no.\ of} \choose \text{constraints}}.   
\end{equation}

\section{Antisymmetric rank-2 tensor \label{AStenssec}}

\subsection{Action}

To illustrate the problems which can arise in a model containing both
constraints and a vacuum manifold, we consider the case of a rank-2
antisymmetric tensor field $B_{ab} = B_{[ab]}$.  In four-dimensional
spacetime, there are two possible invariants that can be constructed
from this field, which we will denote as $X$ and $Y$:
\begin{align} \label{AStens.invardef}
  X &= B^{ab} B_{ab} & Y &= \mathfrak{B}^{ab} B_{ab},
\end{align}
where
\begin{equation}
 \mathfrak{B}^{ab} \equiv \frac{1}{2} \epsilon^{abcd} B_{cd}.
\end{equation}
(Note that $\mathfrak{B}^{ab} \mathfrak{B}_{ab} = - X$.)  We consider
an action of the form
\begin{equation} \label{AStens.lag}
S = \int d^4x \left[ - \frac{1}{12} F_{abc} F^{abc} - V(X,Y) \right],
\end{equation}
where
\begin{equation}
F_{abc} \equiv 3 \partial_{[a} B_{bc]}.
\end{equation}
The Euler-Lagrange equations derived from this action will then be
\begin{equation}
  \frac{1}{2} \partial_c F^{cab} - V_X B^{ab} - V_Y \mathfrak{B}^{ab}
  = 0, \label{AStens.baseEOM}
\end{equation}
where $V_X \equiv \partial V/\partial X$ and $V_Y \equiv \partial
V/\partial Y$. 

% \begin{equation}\label{ASvacmanifold}
%   f(B_{ab}) = B_{ab} B^{ab} - b.
% \end{equation}
% The potential model is
% \begin{equation} \label{AStenspot}
%   \mathcal{L} = - \frac{1}{12} F_{a bc} F^{a bc} - \kappa
%   (B_{a b} B^{a b} -b )^2
% \end{equation}
% while the Lagrange-multiplier model is
% \begin{equation} \label{AStensLM}
% \mathcal{L} =  - \frac{1}{12} F_{a bc} F^{a bc} - \lambda
% (B_{a b} B^{a b} -b ).
% \end{equation} 
% where $\kappa$ is again a constant and $\lambda$ is a Lagrange
% multiplier.  We have also defined the field strength $F_{abc} \equiv
% 3 \partial_{[a} B_{bc]}$.

Since we will be attempting to construct a Hamiltonian for this model,
we will need to perform a 3+1 decomposition. Given a choice of time
coordinate $t = x^0$ on our spacetime, we can decompose the field
$B_{ab}$ into spatial vectors $\vec{P}$ and $\vec{Q}$, corresponding
to its ``electric'' and ``magnetic'' parts respectively:
\begin{align}
P^i &= B^{0i} & Q^i = \frac{1}{2} \epsilon^{ijk} B^{jk},
\end{align}
where $\epsilon^{ijk}$ is the volume element on a constant-$t$
hypersurface in spacetime.  In terms of these, the kinetic term in the
Lagrangians above can be rewritten as
\begin{equation} \label{AStenskinetic}
  -\frac{1}{12} F_{abc} F^{abc} = \frac{1}{2} \left[ \left(\dot{\vec{Q}} - \vec{\nabla} \times
      \vec{P} \right)^2 - (\vec{\nabla} \cdot \vec{Q})^2 \right],
\end{equation}
while the invariants $X$ and $Y$ become
\begin{align}
X &= -2 \vec{P}^2 + 2 \vec{Q}^2, & Y = -4 \vec{P} \cdot \vec{Q}.
\end{align}

For the sake of concreteness in what follows, we will want to have an
explicit form for the potential $V$.  To support Lorentz violation,
the tensor $B_{ab}$ must have a non-zero expectation value which can
couple to other matter fields.  We therefore want to construct a
potential such that there exist solutions to \eqref{AStens.baseEOM}
where $B_{ab}$ is non-zero but constant.  A potential $V$ which is
linear in the invariants $X$ and $Y$ will lead only to solutions where
$B_{ab} = 0$, and we must therefore construct a potential which is
quadratic in the invariants (and hence quartic in the fields):
\begin{equation} \label{AStens.potdef}
  V(B_{ab}) = \frac{1}{2} \kappa_1 X^2 + \kappa_2 XY + \frac{1}{2}
  \kappa_3 Y^2 + \lambda_1 X + \lambda_2 Y, 
\end{equation}
where the $\kappa_i$ and $\lambda_i$ coefficients determine the
``shape'' of the potential.  

Equation \eqref{AStens.baseEOM} will then be satisfied for a constant
tensor field if and only if $V_X = V_Y = 0$, or\footnote{The ``if''
  part of this statement is obvious.  To see the ``only if'' part,
  suppose that $B_{ab} \neq 0$ and $\alpha B_{ab} + \beta
  \mathfrak{B}_{ab} = 0$ for some $\alpha, \beta \in \mathbb{R}$.
  Contracting this equation with $\epsilon^{abcd}$ yields $\alpha
  \mathfrak{B}_{ab} - \beta B_{ab} = 0$, and these two equations
  together imply that $\alpha = \beta = 0$.}
\begin{equation}
  \begin{bmatrix} \kappa_1 & \kappa_2 \\ \kappa_2 &
    \kappa_3 \end{bmatrix} \begin{bmatrix} X \\ Y \end{bmatrix}
  + \begin{bmatrix}  \lambda_1 \\
    \lambda_2 \end{bmatrix} = 0. \label{AStens.vacman}
\end{equation}
Assuming that $\kappa_1 \kappa_3 - \kappa_2^2 \neq 0$, the solutions
to \eqref{AStens.vacman} will be those where $X$ and $Y$ both have a
particular value determined by the $\kappa_i$ and $\lambda_i$
coefficients.  This solution space is the vacuum manifold of our
model; it will be a four-dimensional manifold in the six-dimensional
field space.\footnote{It can be shown that this manifold is
  homeomorphic to $TS^2$, the tangent bundle on the sphere.}  More
generally, the dimension of the vacuum manifold will be four plus the
dimension of the solution space of \eqref{AStens.vacman}.  For
example, for the antisymmetric tensor models discussed in
\cite{Seifert2010a, Seifert2010}, the invariant $Y$ is undetermined,
and thus the vacuum manifold is five-dimensional.

% For the potential \eqref{AStens.potdef} to be bounded below, it is
% necessary that $\kappa_1 \geq 0$, $\kappa_3 \geq 0$, and $\kappa_1
% \kappa_3 - \kappa_2^2 \geq 0$.  To ensure this (and to avoid excessive
% algebraic complications), we will assume in what follows that
% $\kappa_1$ and $\kappa_3$ are strictly positive and $\kappa_2 = 0$,
% unless otherwise stated.

\subsection{Constructing the Hamiltonian}

From the kinetic term in \eqref{AStenskinetic}, we can find the
conjugate momenta for the fields $\vec{P}$ and $\vec{Q}$, the former
of which can be seen to vanish:
\begin{align}
  \vec{\Pi}_P = \frac{\delta \mathcal{L}}{\delta \dot{\vec{P}}} 
  &= 0, 
  & \vec{\Pi}_Q  = \frac{\delta \mathcal{L}}{\delta \dot{\vec{Q}}}
  &= \dot{\vec{Q}} - \vec{\nabla} \times \vec{P}.
\end{align}
Thus, we have three primary constraints, corresponding to the three
components of $\vec{\Phi} \equiv \vec{\Pi}_P = 0$.  The augmented
Hamiltonian will therefore require three Lagrange multipliers, which
we will assemble into a vector $\vec{u}$;  this allows us to write the
augmented Hamiltonian compactly as
\begin{multline}
  \mathcal{H}_A = \vec{\Pi}_Q \cdot \dot{\vec{Q}} - \mathcal{L} +
  \vec{u} \cdot \vec{\Pi}_P \\
  = \frac{1}{2} \vec{\Pi}_Q^2 + \vec{\Pi}_Q \cdot \left( \vec{\nabla}
    \times \vec{P} \right) + \frac{1}{2} (\vec{\nabla} \cdot
  \vec{Q})^2 + V(X,Y) + \vec{u} \cdot \vec{\Pi}_P.
\end{multline}

We must now see whether the primary constraints $\vec{\Phi} = 0$ are
closed under the time-evolution of the system, thereby obtaining
secondary constraints and/or values for the Lagrange multipliers
$\vec{u}$.  The preservation of the primary constraints leads to a set
of secondary constraints $\vec{\Psi}$:
\begin{align} \label{AStens.secconstr} % Notes of 2017-11-03
  0 &= \dot{\vec{\Phi}} = \{ \vec{\Pi}_P, H_A \} \notag \\
    &= \vec{\nabla} \times \vec{\Pi}_Q - \frac{\partial V}{\partial
      \vec{P}} \equiv \vec{\Psi} % \notag \\
    % &= \vec{\nabla} \times \vec{\Pi}_Q + 4 V_X \vec{P} - 4 V_Y \vec{Q}
    %       
\end{align}  % TO CHECK:  SIGN OF CURL TERM HERE
However, preservation of the secondary constraints leads to an
equation involving the unknown Lagrange multipliers $\vec{u}$:
\begin{align}
  0 &= \dot{\Psi_i} 
      = \left\{ (\vec{\nabla} \times \vec{\Pi}_Q)_i - \frac{\partial
      V}{\partial P_i}, H_A \right\} \notag \\
    &= - \left[ \vnab \times \left( \frac{\partial V}{\partial \vec{Q}}
      \right) \right]_i - \frac{\partial^2 V}{\partial P_i \partial P_j}
      u_j \notag \\
    & \qquad \qquad \qquad {} - \frac{\partial^2 V}{\partial P_i
      \partial Q_j} \left[ 
      \vec{\Pi}_Q + \vnab \times \vec{P} \right]_j
    % &= 4 \kappa \left[ -2(B^2 - b) \vec{Q} + 2 \vec{P} \left( \vec{Q}
    %   \cdot (\vec{\Pi}_Q + \vec{\nabla} \times \vec{P} ) \right) +
    %   (B^2 
    %   - b) \vec{u} - 4 \vec{P} (\vec{P} \cdot \vec{u})
    %   \right]
 \label{ASpotudet} 
\end{align}
If we define a vector $\vec{v}$ as
\begin{equation}
  v_i \equiv  \left[ \vnab \times \left( \frac{\partial V}{\partial
        \vec{Q}} 
    \right) \right]_i + \frac{\partial^2 V}{\partial P_i \partial Q_j}
  \left[ 
    \vec{\Pi}_Q + \vnab \times \vec{P} \right]_j
\end{equation}
and a matrix $\mathcal{M}_{ij}$ as
\begin{equation} \label{AStens.Mdef}
  \mathcal{M}_{ij} = \frac{\partial^2 V}{\partial P_i \partial P_j},
\end{equation}
then the equation \eqref{ASpotudet} reduces to the equation
\begin{equation} \label{AStens.constrmat}
\mathcal{M}_{ij} u_j + v_i = 0.
\end{equation}
This equation will determine some or all of the components of
$\vec{u}$; the number of components so determined is equal to the rank
of the matrix $\mathcal{M}$.

What remains is to find an expression for $\mathcal{M}_{ij}$.  Using
the chain rule, it is not hard to show that
\begin{multline}
  \mathcal{M}_{ij} = - 4V_X \delta_{ij} \\
  + 16 \kappa_1 P_i P_j + 32 \kappa_2 P_{(i} Q_{j)}
  + 16 \kappa_3 Q_i Q_j,
\end{multline}
where $V_X = \partial V/\partial X = \kappa_1 X + \kappa_2 Y +
\lambda_1$.  To solve \eqref{AStens.constrmat}, we need to invert
$\mathcal{M}_{ij}$. By taking an ansatz of the form
\begin{subequations} \label{AStens.invmat}
\begin{equation} 
  (\mathcal{M}^{-1})_{jk} = \mathcal{A} \delta_{jk} + \mathcal{B} P_j
  P_k + 2\mathcal{C} P_{(j} Q_{k)} + \mathcal{D} Q_j Q_k
\end{equation}
and requiring that $\mathcal{M}_{ij} (\mathcal{M}^{-1})_{jk} =
\delta_{ik}$, we find that the inverse exists for a generic point in
field space, with
\begin{align}
\mathcal{A} &= -\frac{1}{4 V_X} \\
\mathcal{B} &= \frac{1}{V_X \mathcal{Q}} \left[- \kappa_1 V_X + 4
              \vec{Q}^2 (\kappa_1 \kappa_3 - \kappa_2^2) \right] \\
\mathcal{C} &= \frac{1}{V_X \mathcal{Q}} \left[- \kappa_2 V_X - 4
              (\vec{P} \cdot \vec{Q}) (\kappa_1 \kappa_3 - \kappa_2^2)
              \right] \\ 
\mathcal{D} &= \frac{1}{V_X \mathcal{Q}} \left[- \kappa_3 V_X + 4
              \vec{P}^2 (\kappa_1 \kappa_3 - \kappa_2^2) \right],
\end{align}
where 
\begin{multline}
  \mathcal{Q} \equiv V_X^2 - 4 V_X (\kappa_1 \vec{P}^2 + 2 \kappa_2
  \vec{P} \cdot \vec{Q} + \kappa_3 \vec{Q}^2) \\
  + 16 (\kappa_1 \kappa_3 - \kappa_2^2) \left[ \vec{P}^2 \vec{Q}^2 -
    (\vec{P} \cdot \vec{Q})^2 \right].
\end{multline}
\end{subequations}

At a generic point in field space, this is well-defined, and so we can
invert \eqref{AStens.constrmat} to determine the three Lagrange
multipliers $\vec{u}$ in terms of the other fields.  The overall
Hamiltonian density for the system would then be
\begin{multline} \label{AStens.finalham}
  \mathcal{H}_A = \frac{1}{2} \vec{\Pi}_Q^2 + \vec{\Pi}_Q \cdot \left(
    \vec{\nabla} \times \vec{P} \right) + \frac{1}{2} (\vec{\nabla}
  \cdot \vec{Q})^2 \\
  {} + V(X,Y) - (\Pi_P )_i (\mathcal{M}^{-1})_{ij} v_j.
\end{multline}
Further, if this inverse is well-defined, we can count the number of
degrees of freedom of the theory.  We have six fields ($\vec{P}$ and
$\vec{Q}$), three primary constraints $\vec{\Pi}_P = 0$, and three
secondary constraints given in \eqref{AStens.secconstr}.  Thus, the
number of degrees of freedom for a general point in field space is
\begin{equation}
N_{dof} = 6 - \frac{1}{2} (3 + 3 + 0) = 3.
\end{equation}

It is evident, however, that the inverse matrix \eqref{AStens.invmat}
is \emph{not} well-defined when either $\mathcal{Q}$ or $V_X$ vanish.
This presents a dilemma.  If we start with an initial-data
configuration satisfying the constraints and for which $\mathcal{Q}$
and $V_X$ are non-vanishing, then the Hamiltonian
\eqref{AStens.finalham} becomes singular if the fields ever evolve to
a point where $V_X$ or $\mathcal{Q}$ vanish.  Alternately, one could
construct a Hamiltonian under the assumption that $V_X$ and/or
$\mathcal{Q}$ vanish.  In this case, the matrix $\mathcal{M}_{ij}$
would not be of full rank.  This would leave one or more components of
$\vec{u}$ undetermined in \eqref{AStens.constrmat}; it would also
require that certain components of $\vec{v}$ (those not in the range
of $\mathcal{M}_{ij}$) vanish automatically, leading to additional
constraints.  The iterative constraint-generation procedure described
in Section \ref{HFTsec} would therefore have to continue; assuming
that it did not lead to an inconsistency, the resulting theory would
necessarily have fewer degrees of freedom than the theory constructed
for a generic point in field space.

This ``loss'' of a degree of freedom at certain points in field space
was noted in \cite{Seifert2019} in the context of a vector field model
with an unorthodox kinetic term.  It has also been noted in certain
vector field models in curved spacetime \cite{Isenberg1977,
  Garfinkle2012}.  While these features of those models are troubling,
one could perhaps argue that the singularities of those models occur
at non-generic points in field space that in some sense are
well-separated from ``typical'' field configurations, and therefore
that those models might still be viable.

What makes the singularity in the present case especially vexing,
however, is that we cannot make such an argument.  The Hamiltonian is
singular when $V_X =0$, and \emph{by definition $V_X = 0$ holds for
  all points in the vacuum manifold}.  The above arguments imply that
the evolution between field configurations ``on'' the vacuum manifold
and field configurations ``off'' the vacuum manifold is rather
ill-posed, since Hamilton's equation for $\vec{P}$ is
\begin{equation}
  \frac{d P_i}{dt} = \frac{\delta H_A}{\delta \Pi_{Pi}} =
  - (\mathcal{M}^{-1})_{ij} v_j.
\end{equation}
This casts serious doubt on the viability of such a field theory as a
candidate for dynamical Lorentz symmetry violation.  The field
configurations with $\vec{v} \neq 0$ are generic in field space, both on
and off the vacuum manifold.  Most small perturbations away from the
vacuum manifold would therefore have $\vec{v} \neq 0$ as they evolve
``back towards'' the vacuum manifold.  But since $\mathcal{M}^{-1}$
becomes singular as the fields approach the vacuum manifold, we are
forced to conclude that $d\vec{P}/dt$ will diverge as the fields
evolve back towards the vacuum manifold.

This statement may seem to be at odds with the work of Altschul
\emph{et al.}\ \cite{Altschul2010}.  In that work, the authors
linearized the equations of motion \eqref{AStens.baseEOM} about a
constant background tensor, and explored their properties.  These
properties included the presence of ``massive modes'', in which the
field evolved away from the vacuum manifold.

I believe that the discrepancy here is an issue of the so-called
``linearization stability'' of non-linear differential equations
\cite{Marsden1975}.  In looking at perturbations about a particular
background, it is usually assumed that the solutions to the linearized
equations of motion correspond to ``small solutions'' of the full
non-linear equations of motion.  A theory for which this
correspondence can be drawn is said to be \emph{linearization stable.}
However, not all models have this property; it is entirely possible
that one can find solutions to the linearized equations that do not
correspond to any solutions of the full non-linear equations.  Showing
whether a given set of non-linear differential equations is
linearization stable is a complex question, and a full discussion
would be beyond the scope of this paper; but the present result would
seem to indicate that the antisymmetric tensor evolution equations
\eqref{AStens.baseEOM} are not in fact linearization stable.
% , and that the massive modes found in
% \cite{Altschul2010} do not actually exist in the full non-linear
% model.

More recently, Hernaski has also examined the consequences of
spontaneous Lorentz symmetry violation in the context of an
antisymmetric rank-2 tensor \cite{Hernaski2016}.  That work used
general symmetry considerations to find the most general form of an
effective Lagrangian for the Nambu-Goldstone modes arising from this
sort of spontaneous Lorentz symmetry breaking. While this construction
does not involve the problematic ``massive modes'' of the model, the
fact that the ``low-energy limit'' of an action \eqref{AStens.lag} is
so ill-posed means that it is unclear if Hernaski's effective
Lagrangian could correspond actually correspond to the low-energy
limit of a model involving a fundamental tensor field breaking Lorentz
symmetry.  However, in the construction, Hernaski remained agnostic
about the mechanism by which this vacuum expectation value arose;  and
other mechanisms could possibly still give rise to such an effective
Lagrangian (see Section \ref{disc.sec}.)

\section{General field theories \label{GFT.sec}}

\subsection{Invariants and constraints}

Confronted with this problem, two natural questions arise: why does
this pathology occur, and does it affect other tensor field models?  A
similar pathology was noted in \cite{Seifert2019} for the
``$V$-field'', a model consisting of a vector field $A_a$ governed by
the action
\begin{equation}
  S = \int d^4x \left[ \nabla_a A^b \nabla_b A^a - V(A_a)\right],
\end{equation}
where $V(A^a) = \kappa (A^a A_a - b)^2$.  This Lagrangian can be
integrated by parts to cast it in the alternate form
\begin{equation} \label{Vfieldlag}
  S = \int d^4x \left[ \left(\nabla_a A^a \right)^2 - V(A_a) \right].
\end{equation}
In this form, it is evident that the model has three constraints,
since the velocities of the spatial components $\dot{\vec{A}}$ do not
appear.  However, the $3 \times 3$ matrix $\mathcal{M}_{ij}$ from that
work (defined analogously to \eqref{AStens.Mdef} here) is
\begin{equation}
  \mathcal{M}_{ij} = \frac{\partial^2 V}{\partial A_i \partial A_j} =
  4 \kappa \left[ \delta_{ij} (A_a A^a - b) + 2 A_i A_j \right].
\end{equation}
This can be seen to have a rank of 1 if $A_a A^a = b$ and 3 otherwise.
Since $\mathcal{M}$ has full rank off the vacuum manifold
but has a non-trivial nullspace on the vacuum manifold, the inverse
for $\mathcal{M}$ becomes singular on the vacuum manifold, leading to
the same pathology we found in the antisymmetric tensor case.

In the other two models discussed in \cite{Seifert2019}, however,
the number of constraints is smaller.  For a general kinetic term of
the form
\begin{equation}
  \mathcal{L}_K = c_1 (\nabla_a A_b) (\nabla^a A^b) + c_3 (\nabla_a
  A_b) (\nabla^b A^a)
\end{equation}
and the same potential $V(A_a)$, there is one primary constraint if
$c_1 = - c_3$ (this is the familiar ``Maxwell'' kinetic term), and no
primary constraints if $c_1 \neq - c_3$ and $c_1 \neq 0$.

The pathology therefore seems to depend on the number of primary
constraints in the model.  Specifically, both the $V$-field model and
the antisymmetric tensor model have the property that on the vacuum
manifold, the rank of the matrix $\mathcal{M}$ is less than the number
of primary constraints.  In both cases, the matrix $\mathcal{M}$ is
constructed by taking the second derivatives of the potential $V$ with
respect to the ``constrained fields'': $\vec{P}$ for the antisymmetric
tensor, $\vec{A}$ for the $V$-field.  It is the failure of this
matrix to be full-rank on the vacuum manifold that leads to a singular
Hamiltonian.

It is not hard to show that the rank of any matrix constructed in such
a way is bounded above by the number of invariants used to construct
the potential $V$.  Suppose we have a potential $V(X_1, X_2, \cdots,
X_N)$, where the quantities $X_A$ are in turn functions of some set of
field variables $\psi_\alpha = \{\psi_1, \psi_2, \cdots, \psi_n \}$,
with $n > N$.  The analogous matrix will then be
\begin{equation}
  \mathcal{M}_{\alpha \beta} = \frac{\partial^2 V}{\partial
    \psi_\alpha \partial \psi_\beta} 
\end{equation}
If we imagine diagonalizing this matrix, we can see that the nullspace
of this matrix corresponds to the ``directions'' in field space in
which the potential is flat; in other words, the nullity of
$\mathcal{M}_{\alpha \beta}$ is precisely the dimension of the vacuum
manifold in field space, and the rank of $\mathcal{M}_{\alpha \beta}$
is its codimension in field space.  More explicitly, we can use the
chain rule to rewrite $\mathcal{M}_{\alpha \beta}$ as
\begin{equation}
  \mathcal{M}_{\alpha \beta} = \frac{\partial^2 V}{\partial
    X_A \partial X_B} \frac{\partial X_A}{\partial \psi_\alpha}
  \frac{\partial X_B}{\partial \psi_\beta} + \frac{\partial V}{\partial
    X_A} \frac{\partial^2 X_A}{\partial \psi_\alpha \partial
    \psi_\beta}, \label{generalmdef}
\end{equation} 
where a summation over $A$ and $B$ is understood.  In the vacuum
manifold, the second term in \eqref{generalmdef} will vanish (since
the vacuum manifold, by definition, extremizes $V$ with respect to of
all its arguments.)  The first term, meanwhile, will have a rank of at
most $N$, the number of invariants used to construct $V$.  This
implies that in the vacuum manifold, the rank of $\mathcal{M}_{\alpha
  \beta}$ will be less than $n$.\footnote{In both of the explicit
  models under consideration, the second term in \eqref{generalmdef}
  is of rank $n$ when $V_X \neq 0$; in fact, it works out to be
  proportional to $\delta_{\alpha \beta}$.  This may not occur in a
  more general case.}

Effectively, this means that if the vacuum manifold has ``too many
dimensions'', we risk the rank of this matrix being too small on the
vacuum manifold.  But for a given tensor field, there are only a
limited number of independent Lorentz invariants that can be
constructed from it; and it is possible that a given tensor field may
not have enough invariants to reduce the nullity (and increase the
rank) of $\mathcal{M}_{\alpha \beta}$ sufficiently.

This illustrates why the matrix $\mathcal{M}$ is not of full rank in
either the antisymmetric tensor model or the $V$-field model.  In both
cases, we are constructing the matrix $\mathcal{M}$ by taking the
derivatives of $V$ with respect to three constrained fields: the
``electric vector'' $\vec{P}$ for the antisymmetric tensor, or the
spatial components of $A_a$ for the $V$-field.  But there are only two
invariants that can be constructed out of an antisymmetric tensor
$B_{ab}$, and only one that can be constructed out of a vector field
$A_a$, and so the rank of $\mathcal{M}$ decreases when we are on the
vacuum manifold.

\subsection{Constraint structure \label{gencase.sec}}

Given the above features of the antisymmetric tensor and $V$-field
models, one might conjecture that any model which has ``more
constraints than invariants'' would exhibit a similar pathology.
However, the picture is not so simple.  In particular, the structure
of the constraints was critical to the argument: the problematic
conditions followed from the preservation of the secondary
constraints, and the preservation of the primary constraints did not
determine any of the Lagrange multipliers $u_\alpha$ in any way.  In
this section, I will therefore proceed through the constraint algebra
for a more general field theory to see under which circumstances the
simple picture of the pathology arising from ``more constraints than
invariants'' might hold.

Suppose we consider a field theory in terms of some set of fields
$\psi_\alpha$ ($1 \leq \alpha \leq n$) whose dynamics are given by a
Lagrangian that is quadratic in these fields' derivatives, both
spatial and temporal.  Suppose, further, that the ``kinetic terms''
$\mathcal{L}_K$ of the Lagrangian depend only on these derivatives, so
that we have
\begin{multline}
  \mathcal{L}_K = \frac{1}{2} \sum_{\alpha, \beta \leq n} \left[
    \mathcal{P}_{\alpha \beta} \dot{\psi}_\alpha \dot{\psi}_\beta + 2
    \mathcal{Q}_{\alpha i \beta} \dot{\psi}_\alpha \nabla_i \psi_\beta
  \right. \\
  \left. {} + \mathcal{R}_{i \alpha j \beta} \nabla_i \psi_\alpha
    \nabla_j \psi_\beta \right] \label{genftarbform}
\end{multline}
The quantities $\mathcal{P}_{\alpha \beta}$, $\mathcal{Q}_{\alpha i
  \beta}$, and $\mathcal{R}_{i \alpha j \beta}$ are numerical
coefficients, independent of the fields and of spacetime
coordinates. Here and in what follows, we will need to explicitly
write out the summations over field indices; repeated indices should
\emph{not} be assumed to be summed if the summation is not stated
explicitly.  However, the summations over the spatial indicies $i$ and
$j$ will remain implicit.

Via various field and coefficient redefinitions, it can be shown (see
Appendix \ref{lagredef.sec}) that a set of kinetic terms of this form
can always be rewritten in the form
\begin{widetext}
\begin{multline}
  \mathcal{L}_K = \frac{1}{2} \sum_{\alpha, \beta \leq m}
  \mathcal{P}_{\alpha \beta} \left( \dot{\psi}_\alpha + \sum_{\gamma
      \leq n} \mathcal{S}_{\alpha i \gamma} \nabla_i \psi_\gamma
  \right) \left( \dot{\psi}_\beta + \sum_{\delta
      \leq n} \mathcal{S}_{\beta i \delta} \nabla_i \psi_\delta
  \right) \\
  {} + \sum_{\alpha, \beta > m} \mathcal{Q}_{\alpha i \beta}
  \dot{\psi}_\alpha \nabla_i \psi_\beta + \frac{1}{2} \sum_{\alpha,
    \beta \leq n} \mathcal{R}_{i \alpha j
    \beta} \nabla_i \psi_\alpha \nabla_j \psi_\beta. \label{canonlag}
\end{multline}
\end{widetext}
where $\mathcal{P}_{\alpha \beta}$ is a non-degenerate diagonal matrix
and $m \leq n$.  The advantage of this form is that it is
particularly simple to identify the primary constraints.  For $\alpha
\leq m$, we have
\begin{equation}
  \pi_\alpha \equiv \frac{\partial \mathcal{L}_K}{\partial
    \dot{\psi}_\alpha} = \mathcal{P}_{\alpha
    \alpha} \left( \dot{\psi}_\alpha + \sum_{\delta
      \leq n} \mathcal{S}_{\alpha i \delta} \nabla_i \psi_\delta
  \right),
\end{equation}
which can be easily inverted to find the velocities in terms of the
momenta and derivatives.  For $\alpha > m$, meanwhile, we have
\begin{equation} 
  \pi_\alpha = \sum_{\beta > m} \mathcal{Q}_{\alpha i \beta} \nabla_i
  \psi_\beta,
\end{equation}
which can easily be seen to be a constraint equation:
\begin{equation}
  \label{canonprimconst}
  \Phi_\alpha \equiv \pi_\alpha -
 \sum_{\beta > m} \mathcal{Q}_{\alpha i \beta} \nabla_i
  \psi_\beta = 0. 
\end{equation}
The number of primary constraints in this model is therefore $n - m$.

Let us now suppose that the full Lagrangian density of the model is of
the form
\begin{equation}
  \mathcal{L} = \mathcal{L}_K - V(\psi_\alpha) 
\end{equation}
with $\mathcal{L}_K$ of the form given in \eqref{canonlag} and
$V(\psi_\alpha)$ a potential that does not depend on any field
derivatives.  Then the base Hamiltonian density of this model will be 
\begin{multline}
  \label{canonham}
  \mathcal{H}_0 = \frac{1}{2} \sum_{\alpha, \beta \leq m}
  \mathcal{P}_{\alpha \beta}^{-1} \pi_\alpha \pi_\beta - \sum_{\alpha
    \leq m} \sum_{\beta \leq n} \pi_\alpha \mathcal{S}_{\alpha i
    \beta} \nabla_i \psi_\beta \\ {} - \frac{1}{2} \sum_{\alpha,
    \beta \leq n} \mathcal{R}_{i \alpha j
    \beta} \nabla_i \psi_\alpha \nabla_j \psi_\beta + V(\psi_\alpha),
\end{multline}
The augmented Hamiltonian density can then be obtained by adding
\begin{equation}
  \label{canonhlm}
  \mathcal{H}_{LM} = \sum_{\alpha > m} u_\alpha \left( \pi_\alpha -
    \sum_{\beta > m} \mathcal{Q}_{\alpha i \beta} \nabla_i \psi_\beta \right)
\end{equation}
to \eqref{canonham}.

The first requirement we will need to impose to reproduce the
pathology found in the previous section is to require that the
coefficients $\mathcal{Q}_{\alpha i \beta}$ in \eqref{canonlag} vanish
for $\alpha, \beta > m$.  To see this, note that the time-evolution of
a primary constraint $\Phi_\alpha$ will be given by
\begin{equation}
  \dot{\Phi}_\alpha = \{ \Phi_\alpha, H_0 \} + \{ \Phi_\alpha, H_{LM}
  \}, \label{canonprimpres}
\end{equation}
where $H_0 \equiv \int d^3x \mathcal{H}_0$ and $H_{LM} \equiv \int
d^3x \mathcal{H}_{LM}$.  The latter Poisson bracket can be evaluated
to be
\begin{equation} \label{canonfirstclasscons}
  \{ \Phi_\alpha, H_{LM} \} = - \sum_{\beta > m} \mathcal{Q}_{\alpha i
    \beta} \nabla_i u_\beta.
\end{equation}
As noted above, the pathology in the previous section arises from the
preservation of the secondary constraints, not the primary
constraints.  If the Lagrange multipliers enter at this stage, then
the preservation of the \emph{primary} constraints will at least
partially determine them, and the chain of logic will diverge at this
stage.  Thus, for a model to follow the same logical chain, we must
have these terms vanishing for all $\alpha$; and so we must have
$\mathcal{Q}_{\alpha i \beta} = 0$.  In the language of Dirac, this
means that the primary constraints are all first-class, since they all
mutually commute with each other. Both the antisymmetric tensor model
and the $V$-field model have this property.

Making this assumption, the secondary constraints $\Psi_\alpha \equiv
\{ \Phi_\alpha, H_0 \}$ can be calculated to be
\begin{equation}
  \Psi_\alpha = - \frac{\partial V}{\partial \psi_\alpha} -
  \sum_{\beta \leq n} \left( \mathcal{S}_{\alpha i
      \beta} \nabla_i \pi_\beta + \mathcal{R}_{i \alpha j \beta}
    \nabla_i \nabla_j \psi_\beta \right).
\end{equation}
These secondary constraints will in turn need to be preserved, i.e.,
$\dot{\Psi}_\alpha = \{ \Psi_\alpha, H_A \} = 0$.  In the pathological
cases described above, the Lagrange multipliers entered into the
analogous equation.  To see where they enter here, we can calculate
the Poisson bracket $\{ \Phi_\alpha, H_A \}$;  we obtain
\begin{equation}
  \dot{\Psi}_\alpha = v_\alpha - \sum_{\beta > m} \left(
    \frac{\partial^2 V}{\partial \psi_\alpha \partial \psi_\beta} u_\beta +
    \mathcal{R}_{i \alpha j \beta} \nabla_i \nabla_j u_\beta
  \right), \label{canonsecpres}
\end{equation}
where $v_\alpha$ represents all terms that do not depend on the
Lagrange multipliers $u_\alpha$.  

The first term in the parentheses in \eqref{canonsecpres} is the one
that caused the pathology in the cases of the antisymmetric tensor and
the $V$-field;  the rank of the matrix 
\begin{equation}
  \frac{\partial^2 V}{\partial \psi_\alpha \partial \psi_\beta} =
  \mathcal{M}_{\alpha \beta}
\end{equation}
was different on the vacuum manifold than on a general point in field
space.  However, we can see from the above that in a more general
model, it is possible for more of the Lagrange multipliers than one
would expect from the rank of $\mathcal{M}_{\alpha \beta}$ to be
determined by the preservation of the secondary constraints; the
derivative terms in \eqref{canonsecpres} can also help to determine
the Lagrange multipliers.  For a model to have the vacuum manifold
pathology described in the previous sections, it is sufficient for
these derivative terms to vanish; in other words, $\mathcal{R}_{i
  \alpha j \beta} = 0$ for $\alpha, \beta > m$.\footnote{Note that we
  can take $\mathcal{R}_{i \alpha j \beta}$ to be symmetric under
  exchange of $i$ and $j$, since $\mathcal{R}_{i \alpha j \beta}
  \nabla_i \psi_\alpha \nabla_j \psi_\beta = \mathcal{R}_{i \alpha j
    \beta} \nabla_j \psi_\alpha \nabla_ji \psi_\beta$ up to total
  derivatives.}

In summary, then: any Lagrangian whose kinetic terms are of the form
\eqref{canonlag} (or can be put into this form) will suffer from the
vacuum manifold pathology described above if the coefficients
$\mathcal{Q}_{\alpha i \beta}$ and $\mathcal{R}_{i \alpha j \beta} =
0$ when both $\alpha$ and $\beta$ are greater than $m$, and when the
potential $V$ is constructed from fewer than $n-m$ field quantities
depending on the fields $\psi_\alpha$ (with $\alpha > m$).  Both the
antisymmetric tensor model and the $V$-field model satisfy these
criteria.  The kinetic term \eqref{AStenskinetic} is of the form
\eqref{canonlag}; a term corresponding to the $\mathcal{R}_{i \alpha j
  \beta}$ term does exist in the kinetic terms (specifically, the term
$(\vnab \cdot \vec{Q})^2$), but it only involves the ``unconstrained''
fields $\vec{Q}$ and not the ``constrained'' fields $\vec{P}$.
Similarly, the kinetic term for the $V$-field Lagrangian
\eqref{Vfieldlag}, when decomposed into its time and space components,
is
\begin{equation}
  \mathcal{L}_K = (\dot{A}_0 - \vec{\nabla} \cdot \vec{A})^2
\end{equation}
which does not contain any terms corresponding to the second or third
summations in \eqref{canonlag} at all.

The above-listed conditions appear to be sufficient for this
pathology, but they may not be necessary.  It is entirely possible
that a model for which some of the primary constraints were
second-class, or for which the ``constrained fields'' appeared with
spatial derivatives could still have a Hamiltonian which became
singular on the vacuum manifold.  However, in either case, the
Lagrange multipliers cannot be solved for algebraically, and so the
analysis of the Hamiltonian would not be nearly as straightforward.
It is also unclear whether any physically well-motivated models exist
with these features.

% It would be simple to modify the kinetic terms of a model with this
% vacuum manifold pathology in order to make it less problematic.  For
% example, one could modify the antisymmetric two-tensor model to
% eliminate the vacuum manifold pathology by adding the term $(\nabla_i
% P_j) (\nabla_i P_j)$ to the Lagrangian.  This would explicitly break
% Lorentz symmetry, but would lead to all of the Lagrange multipliers
% being automatically determined: the equation \eqref{AStens.constrmat}
% would become
% \begin{equation}
%   \nabla^2 u_i + \mathcal{M}_{ij} u_j + v_i = 0,
% \end{equation}
% and so all three Lagrange multipliers would be determined (up to
% boundary conditions) both on and off the vacuum manifold, eliminating
% the problematic singular Hamiltonian of the original model.  It may be
% the case that for a model with underlying Lorentz symmetry, the
% implicit symmetry between spatial and temporal derivatives force the
% $\mathcal{Q}_{\alpha i \beta}$ and $\mathcal{R}_{i \alpha j \beta}$
% coefficients to automatically vanish.  If so, one could simply
% identify models with pathological vacuum manifolds by simply counting
% the number of constraints and the number of Lorentz invariants used in
% the model.  However, the evidence for this conjecture is slim; the
% three vector models from \cite{Seifert2019} and the antisymmetric
% tensor model all obey this conjecture, but four is a very small number
% among all the possible field theories one could conceivably write down.

\section{Discussion \label{disc.sec}}

We have shown that a model which spontaneously breaks Lorentz symmetry
may, unless constructed with care, have pathological evolution that is
not immediately evident at the level of the Lagrangian.  This result
has serious implications for the construction of such models, and one
would like to know how to evade these pathologies.

Assuming that the model under consideration can be written in the form
\eqref{canonlag}, with $\mathcal{Q}_{\alpha i \beta} = 0$ and
$\mathcal{R}_{i \alpha j \beta} = 0$ when $\alpha, \beta > m$, then
the only way to avoid the pathology is to ensure that the rank of the
matrix $\mathcal{M}_{\alpha \beta}$ is sufficiently high.  This may
not always be possible.  For example, for the case of an antisymmetric
2-tensor $B_{ab}$, there are only two independent Lorentz invariants
that can be constructed from it, namely $X$ and $Y$ as defined in
\eqref{AStens.invardef}.  Since the kinetic term used in
\eqref{AStens.lag} gives rise to three constraints, it is simply not
possible to write down a potential for $B_{ab}$ that does not give
rise to pathological evolution.  One could possibly change the kinetic
term so that the model had no more than two primary constraints;
however, this could very well give rise to instabilities, as discussed
in the Introduction.  

For tensor fields of higher rank or different symmetry type, a similar
set of considerations would have to come into play.  As an example,
Kosteleck\'y and Potting's linear cardinal gravity model
\cite{Kostelecky2009} involves a \emph{symmetric} rank-two tensor
field $C_{ab}$ in a flat background with a potential $V(C_{ab})$ and
the standard kinetic terms for a massless spin-2 field:
\begin{multline}
  \mathcal{L}_K = -\frac{1}{4} \left[ \nabla_c C_{ab} \nabla^c
    C^{ab} - \nabla_a C \nabla^a C \right. \\
  \left. + 2 \nabla_a C \nabla_b C^{ab} - 2
    \nabla_b C^{ab} \nabla_c C_a {}^c \right]
\end{multline}
Performing a 3+1 decomposition, we find that the time derivatives of
$C_{00}$ and $C_{0i}$ do not appear in this Lagrangian, and thus this
model will have four primary constraints.  Moreover, there are no
cross-couplings between the derivatives (either spatial or temporal)
of $C_{00}$ and $C_{0i}$; this means that the appropriate
$\mathcal{Q}_{\alpha i beta}$ and $\mathcal{R}_{i \alpha j \beta}$
coefficients vanish in order for the general result of Section
\ref{gencase.sec} to hold.

Given these features of linear cardinal gravity, one might be
concerned that it runs the risk of suffering from the same
vacuum-manifold pathology as the antisymmetric rank-2 tensor field.
However, there is one important distinction:  while there are only two
independent invariants one can construct from an antisymmetric tensor
field ($X$ and $Y$), there are four independent invariants that can be
constructed from a symmetric rank-2 tensor field:
\begin{subequations}
\begin{align}
  X_1 &= C_a {}^a \\
  X_2 &= C_a {}^b C_b {}^a\\
  X_3 &= C_a {}^b C_b {}^c C_c {}^a\\
  X_4 &= C_a {}^b C_b {}^c C_c {}^d C_d {}^a 
\end{align}
\end{subequations}
Thus, there appear to be ``just enough'' invariants for a cardinal
gravity model to avoid the vacuum-manifold pathology discussed in this
work, so long as the potential has non-trivial dependencies on all
four of these invariants.  It must be emphasized, however, that having
``no more constraints than invariants'' is a necessary, not
sufficient, condition to obtain a non-singular Hamiltonian that can be
extended throughout all of field space;  it is possible that other,
more subtle pathologies occur in such a model.

In the face of these difficulties, it is important to note that there
are other methods by which models can include ``naturally non-zero''
fields.  One could, for example, postulate that $B_{ab}$ is not a
fundamental field but is instead a function of some other fundamental
field which gains a vacuum expectation value.  For example, a recent
work by Assun\c{c}\~{a}o \emph{et al.}  \cite{Assuncao2019} introduced
a model in which $B_{ab}$ is a spinor condensate.  It is not
immediately clear whether the present results could be generalized to
such models.

Another method to avoid these difficulties would be to simply
constrain the field to be non-zero via the use of a Lagrange
multiplier $\lambda$:
\begin{equation} \label{AStens.LMaction}
  S = \int d^4 x \left[ -
    \frac{1}{12} F_{abc} F^{abc} - \lambda (X - b) \right],
\end{equation}
where $X$ is defined as in \eqref{AStens.invardef} and $b \neq 0$ is a
constant.  The equation of motion for $\lambda$ is then simply $X =
b$, and thus the tensor field $B_{ab}$ would be non-zero ``in
vacuum''.\footnote{It is worth noting here that Hernaski's effective
  low-energy Lagrangian \cite{Hernaski2016} could equally well arise
  from a ``fundamental'' Lagrangian of this type, rather than from a
  Lagrangian involving a potential for the fundamental field
  $B_{ab}$.} 

This method introduces a new field to the model as well as a new
primary constraint; more importantly, it also changes the algebra of
the primary constraints in important ways \cite{Seifert2019}.  If we
perform Dirac-Bergmann analysis on this Lagrangian, the augmented
Hamiltonian density can be shown to be
\begin{multline}
  \mathcal{H}_A = \frac{1}{2} \vec{\Pi}_Q^2 + \vec{\Pi}_Q \cdot \left(
    \vec{\nabla} 
    \times \vec{P} \right) + \frac{1}{2} (\vec{\nabla} \cdot
  \vec{Q})^2 \\+ \lambda (X- b) + \vec{u} \cdot \vec{\Pi}_P +
  u_{\lambda} \varpi.
\end{multline}
Here, $\varpi$ is the conjugate momentum to $\lambda$; it is
identically zero, and so the equation $\varpi = 0$ is enforced by a
fourth Lagrange multiplier $u_\lambda$ (in addition to the three
Lagrange multipliers $\vec{u}$ enforcing the constraint $\vec{\Pi}_P =
0$.)  If the primary constraints are to be preserved, their Poisson
brackets with the augmented Hamiltonian $H_A$ must vanish.  The
secondary constraints are found to be
% Notes of Jul 2 2017 \& Jan 30 2019
\begin{equation}
  \vec{\Psi} \equiv 4 \lambda \vec{P} - \vnab \times \vec{\Pi}_Q
\end{equation}
and
\begin{equation}
  \Psi \equiv - (X - b).
\end{equation}
These secondary constraints must in turn be preserved under time
evolution; after some algebra, it can be shown that
\begin{equation}
  \dot{\vec{\Psi}} = 4 \left[ u_\lambda \vec{P} + \lambda \vec{u} + \vnab
    \times (\lambda \vec{Q}) \right]
\end{equation}
and
\begin{equation}
  \dot{\Psi} = 4 \left[ \vec{P} \cdot \vec{u} - 4 \vec{Q} \cdot \left(
      \vec{\Pi}_Q 
      + \vnab \times \vec{P} \right) \right].
\end{equation}
These equations uniquely determine the Lagrange multipliers $\vec{u}$
and $u_\lambda$, so long as $\vec{P} \neq 0$ and $\lambda \neq 0$:
\begin{align}
  u_\lambda &= - \frac{1}{P^2} \left[ \vec{P} \cdot \vnab \times
  (\lambda \vec{Q}) + \lambda \vec{Q} \cdot \left(
  \vec{\Pi}_Q 
  + \vnab \times \vec{P} \right) \right] \\
  \vec{u} &= -\frac{1}{\lambda} \left[ u_\lambda \vec{P} + \vnab \times
  (\lambda \vec{Q}) \right].
\end{align}
For generic points in the vacuum manifold, the Hamiltonian is
non-singular, and the vacuum manifold pathology does not arise.  In
some sense, this is not a surprise: the pathology in the potential
model arises when the field evolves onto or off of the vacuum
manifold, but the field is ``stuck'' on the vacuum manifold in the
Lagrange-multiplier model.

It must be said that some authors (myself included) find the use of
Lagrange multipliers to be somewhat inelegant.  This prejudice arises
from classical particle mechanics, where a constrained model can often
be viewed as a limit: one imagines a model where a potential energy is
minimized on the constraint surface, and then takes the limit of this
model as the potential becomes infinitely strong.  From this
perspective, Lagrange multipliers are just an \emph{ad hoc}
approximation to a more fundamental theory.  However, the situation is
a lot more nuanced than that in classical field theory, particularly
in the presence of models with primary constraints \cite{Seifert2019}.
The present work shows that the use of Lagrange multipliers and/or
composite fields may be unavoidable if one wants to model Lorentz
symmetry violation with a tensor field, particularly one of higher
rank; and that Lagrange-multiplier models may not be relatable to a
``more fundamental'' potential model at all.

\acknowledgments

The author would like to thank J.\ Tasson, B.\ Altschul, V.\ A.\
Kosteleck\'{y}, Q.\ Bailey, and D.\ Garfinkle for discussions during
the preparation of this work.  This research was supported in part by
Perimeter Institute for Theoretical Physics.  Research at Perimeter
Institute is supported by the Government of Canada through Industry
Canada and by the Province of Ontario through the Ministry of Economic
Development \& Innovation.

\appendix

\section{Canonical form of field theory Lagrangian \label{lagredef.sec}}

Consider a Lagrangian whose kinetic term is of the form
\eqref{genftarbform} (reproduced here):
\begin{multline*}
  \mathcal{L}_K = \frac{1}{2} \sum_{\alpha, \beta \leq n} \left[
    \mathcal{P}_{\alpha \beta} \dot{\psi}_\alpha \dot{\psi}_\beta + 2
    \mathcal{Q}_{\alpha i \beta} \dot{\psi}_\alpha \nabla_i \psi_\beta
  \right. \\
  \left. {} + \mathcal{R}_{i \alpha j \beta} \nabla_i \psi_\alpha
    \nabla_j \psi_\beta \right] 
\end{multline*}
In Section \ref{gencase.sec}, it was stated that a set of kinetic
terms of this form can always be put into the form \eqref{canonlag}.
This appendix describes how this may be accomplished.

From the form of \eqref{genftarbform}, it is fairly evident that we can
take $\mathcal{P}_{\alpha \beta}$ to be symmetric under the exchange
of $\alpha$ and $\beta$.  Less evident, but equally important, is that
$\mathcal{Q}_{\alpha i \beta}$ can also be taken to be symmetric under
this exchange.  The antisymmetric part of this array (when contracted
with $\dot{\psi}_\alpha \nabla_i \psi_\beta$) can be expressed in
terms of total derivatives:
\begin{multline}
  ( \mathcal{Q}_{\alpha i \beta} - \mathcal{Q}_{\beta i \alpha} )
  \dot{\psi}_\alpha \nabla_i \psi_\beta = \mathcal{Q}_{\alpha i \beta}
  \left( \dot{\psi}_\alpha \nabla_i \psi_\beta -
    \dot{\psi}_\beta \nabla_i \psi_\alpha \right) \\
  = \nabla_i \left( \mathcal{Q}_{\alpha i \beta} \dot{\psi}_\alpha
    \psi_\beta \right) - \frac{\partial}{\partial t} \left(
    \mathcal{Q}_{\alpha i \beta} (\nabla_i \psi_\alpha) \psi_\beta
  \right).
\end{multline}

With this in mind, the procedure for putting the kinetic terms in the
form \eqref{canonlag} is as follows:
\begin{enumerate}
\item Since $\mathcal{P}_{\alpha \beta}$ can be taken to be real and
  symmetric, we can redefine the fields $\psi_{\alpha}$ (via an
  invertible linear transformation) so that the matrix
  $\mathcal{P}_{\alpha \beta}$ becomes diagonal.  Moreover, we can
  reorder these fields so that $\mathcal{P}_{\alpha \alpha} \neq 0$
  for all $\alpha \leq m$, and $\mathcal{P}_{\alpha \alpha} = 0$ for
  $\alpha > m$.  The kinetic terms of the Lagrangian then become
  \begin{multline}
    \mathcal{L}_K = \frac{1}{2} \sum_{\alpha, \beta \leq m}
    \mathcal{P}_{\alpha \beta} \dot{\psi}_\alpha \dot{\psi}_\beta + 
    \sum_{\alpha, \beta \leq n} \mathcal{Q}_{\alpha i \beta}
    \dot{\psi}_\alpha \nabla_i \psi_\beta \\ {} + \frac{1}{2} \sum_{\alpha,
      \beta \leq n} \mathcal{R}_{i \alpha j
      \beta} \nabla_i \psi_\alpha \nabla_j \psi_\beta,
    \label{lagredef1}
  \end{multline}
  where the arrays $\mathcal{Q}_{\alpha i \beta}$ and $\mathcal{R}_{i
    \alpha j \beta}$ have been redefined after the field
  transformation.  

\item By splitting the second sum in \eqref{lagredef1} into three
  parts, integrating by parts, and applying the symmetry of
  $\mathcal{Q}_{\alpha i \beta}$, we can show that
  \begin{align}  
    & \sum_{\alpha, \beta \leq n} \mathcal{Q}_{\alpha i \beta}
    \dot{\psi}_\alpha \nabla_i \psi_\beta \notag\\
    & = \sum_{\alpha \leq m}
      \sum_{\beta \leq n} \mathcal{Q}_{\alpha i \beta} \dot{\psi}_\alpha
      \nabla_i \psi_\beta + \sum_{\alpha \leq m} \sum_{\beta > m}
      \mathcal{Q}_{\alpha i \beta} \dot{\psi}_\alpha \nabla_i
      \psi_\beta \notag  \\
    & \qquad {} + \sum_{m< \alpha, \beta \leq n} \mathcal{Q}_{\alpha i \beta}
      \dot{\psi}_\alpha \nabla_i \psi_\beta \notag \\
    &= \sum_{\alpha \leq m} \sum_{\beta \leq n}
    \tilde{\mathcal{Q}}_{\alpha i \beta} \dot{\psi}_\alpha \nabla_i
    \psi_\beta + \sum_{m< \alpha, \beta \leq n} \mathcal{Q}_{\alpha i
      \beta} \dot{\psi}_\alpha \nabla_i \psi_\beta,
  \end{align}
  where
  \begin{equation}
    \tilde{\mathcal{Q}}_{\alpha i \beta} = \begin{cases}
      \mathcal{Q}_{\alpha i \beta} & \alpha \leq m, \beta \leq m \\
      2 \mathcal{Q}_{\alpha i \beta} & \alpha \leq m, \beta > m
    \end{cases}
  \end{equation}
  
\item Since $\mathcal{P}_{\alpha \beta}$ is a diagonal, nondegenerate
  matrix for $\alpha, \beta \leq m$, it has an inverse.  By then
  defining
  \begin{equation}
    \mathcal{S}_{\alpha i \beta} \equiv \sum_{\gamma \leq m}
    \mathcal{P}^{-1}_{\alpha \gamma} \tilde{\mathcal{Q}}_{\gamma i
      \beta},
  \end{equation}
  we can then show that the first two summations in \eqref{lagredef1}
  are equal to
  \begin{widetext}
  \begin{multline}
    \frac{1}{2} \sum_{\alpha, \beta \leq m}
    \mathcal{P}_{\alpha \beta} \dot{\psi}_\alpha \dot{\psi}_\beta + 
    \sum_{\alpha, \beta \leq n} \mathcal{Q}_{\alpha i \beta}
    \dot{\psi}_\alpha \nabla_i \psi_\beta \\
    = \frac{1}{2} \sum_{\alpha, \beta \leq m}
    \mathcal{P}_{\alpha \beta} \left( \dot{\psi}_\alpha + \sum_{\gamma
        \leq n} \mathcal{S}_{\alpha i \gamma} \nabla_i \psi_\gamma
    \right) \left( \dot{\psi}_\beta + \sum_{\delta
        \leq n} \mathcal{S}_{\beta i \delta} \nabla_i \psi_\delta
    \right) \\
    {} + \sum_{\alpha, \beta > m} \mathcal{Q}_{\alpha i \beta}
    \dot{\psi}_\alpha \nabla_i \psi_\beta 
    - \frac{1}{2} \sum_{\alpha, \beta \leq m} \sum_{\delta, \gamma
      \leq n} \mathcal{P}_{\alpha \beta} \mathcal{S}_{\alpha i \gamma}
    \mathcal{S}_{\beta i \delta} \nabla_i \psi_\gamma \nabla_j
    \psi_\delta.
    \label{lagredef2}
  \end{multline}
  Effectively, what have done here is to simply ``complete the
  square'' of the first two summations in \eqref{lagredef1}.

\item The last sum in \eqref{lagredef2} can then be absorbed into the
  last term in \eqref{lagredef1}, yielding a set of kinetic terms of
  the desired form:
  \begin{multline}
    \mathcal{L}_K = \frac{1}{2} \sum_{\alpha, \beta \leq m}
    \mathcal{P}_{\alpha \beta} \left( \dot{\psi}_\alpha + \sum_{\gamma
        \leq n} \mathcal{S}_{\alpha i \gamma} \nabla_i \psi_\gamma
    \right) \left( \dot{\psi}_\beta + \sum_{\delta
        \leq n} \mathcal{S}_{\beta i \delta} \nabla_i \psi_\delta
    \right) \\
    {} + \sum_{\alpha, \beta > m} \mathcal{Q}_{\alpha i \beta}
    \dot{\psi}_\alpha \nabla_i \psi_\beta + \frac{1}{2} \sum_{\alpha,
      \beta \leq n} \mathcal{R}_{i \alpha j
      \beta} \nabla_i \psi_\alpha \nabla_j \psi_\beta.
  \end{multline}
  
\end{widetext}
  
\end{enumerate}

\bibliographystyle{unsrt}
\bibliography{vector_tensor_dofs}

\end{document}